\documentclass[11pt,a4paper]{article}

\usepackage{graphicx}
\usepackage{amsmath,amsfonts,amsthm}
\renewcommand{\baselinestretch}{1.3} 

\renewcommand{\theequation}{\thesection.\arabic{equation}}

\newcommand{\w}{\omega}
\newcommand{\ph}{\varphi}
\newcommand{\vecr}{\mathbf{r}}

\begin{document}
\pagenumbering{roman} \setcounter{page}{0} 
\title{Four-dimensional polymer collapse II:\\ Interacting self-avoiding trails}
\author{T. Prellberg\dag\thanks{{\tt {\rm
email:}\ thomas.prellberg@tu-clausthal.de}, Phone:\ +49-5323 72 2085,
Fax:\ +49-5323 72 3799} \, and A. L. Owczarek\ddag \\
        \dag Institut f\"ur Theoretische Physik,\\
        Technische Universit\"at Clausthal,\\ Arnold Sommerfeld Stra\ss e
6,\\ D-38678 Clausthal-Zellerfeld,\\ Germany\\
\ddag Department of Mathematics and Statistics,\\
         The University of Melbourne,\\
         3010, Victoria,\\
	 Australia.
 }

\date{April 9, 2001
}

\maketitle 
 
\begin{abstract} 

We have simulated four-dimensional interacting self-avoiding trails
(ISAT) on the hyper-cubic lattice with standard interactions at a wide
range of temperatures up to length $4096$ and at some temperatures up
to length $16384$. The results confirm the earlier prediction (using
data from a non-standard model at a single temperature) of a collapse
phase transition occurring at finite temperature. Moreover they are in
accord with the phenomenological theory originally proposed by
Lifshitz, Grosberg and Khokhlov in three dimensions and recently given
new impetus by its use in the description of simulational results for
four-dimensional interacting self-avoiding walks (ISAW).  In fact, we
argue that the available data is consistent with the conclusion that
the collapse transitions of ISAT and ISAW lie in the same universality
class, in contradiction with long-standing predictions.  We deduce
that there exists a pseudo-first order transition for ISAT in four
dimensions at finite lengths while the thermodynamic limit is
described by the standard polymer mean-field theory (giving a
second-order transition), in contradiction to the prediction that the
upper critical dimension for ISAT is $d_u=4$.

\vspace{1cm} 
 

\end{abstract} 

\vfill

\newpage

\pagenumbering{arabic}

\section{Introduction} 
\setcounter{page}{1}

The geometric collapse of polymers in dilute solution, being one of
the fundamental and most well studied phase transitions in statistical
mechanics attracts continuing interest from both the theoretician
attempting to understand the subtleties inherent in this phenomenon and
the experimentalist wanting a firm foundation for interpreting their
results on more complicated systems. One approach has been the use of
lattice models of various types of \emph{self-avoiding} paths to
describe the geometry of linear polymers with added local potentials
to broadly account for the complex solvent-polymer and polymer-polymer
interactions. The canonical lattice path utilised in this regard has
been the self-avoiding walk (SAW) as it explicitly demonstrates the
``excluded volume'' expected of physical polymers. However, much of
our understanding of three-dimensional polymer collapse
\cite{gennes1975a-a,stephen1975a-a,duplantier1982a-a} has come from
calculations using the continuum Edwards model
\cite{duplantier1986b-a,duplantier1987d-a} which is based upon intersecting
random paths, rather than interacting self-avoiding walks
(ISAW). Hence it is of interest to study lattice models closer in
nature to the Edwards model. One such model is based on the lattice
paths known as self-avoiding trails (SAT)
\cite{malakis1976a-a,shapir1984a-a,lim1988a-a,chang1988a-a} rather
than self-avoiding walks. The type of interaction considered (contact
versus nearest-neighbour for ISAW) is also more closely analogous to
the Edwards model. It is also worth noting that the upper critical
dimension for polymer collapse is expected to be $d_u=3$, and therefore many
subtle logarithmic corrections are predicted to occur in three
dimensions. As is often the case in the study of critical phenomena
our understanding is enhanced by considering the dimension of the
problem as a parameter that can be varied. Any discrepancies between
the behaviour of competing models in four dimensions for example would
impinge on our interpretation of results in three dimensions.

Self-avoiding trails (SAT) or trails for short are paths on a lattice
which have no two steps on the same bond of that lattice but may
occupy the same site.  This restriction is sometimes referred to as
bond-avoiding, in contrast to self-avoiding walks (SAW) which are site-avoiding 
(that is, no two vertices of the walk may occupy the same
site on the lattice).  Clearly walks are, by default, also
bond-avoiding.  Trails possess an excluded volume effect and it is
fairly well established that SAT and SAW are in the same universality class
\cite{guttmann1985b-a,guttmann1985c-a} which describes good solvent polymers.  
(For a more recent discussion of the subtle differences between walks
and trails regarding corrections to scaling see \cite{guim1997a-a}.)
It has been shown \cite{shapir1984a-a} that there should exist a
collapse transition when contact attraction is added to the trail
model: this model of polymer collapse is known as interacting
self-avoiding trails (ISAT). Moreover, Shapir and Oono
\cite{shapir1984a-a} have argued that this point should be tricritical
in nature, as it is at the ISAW collapse point.  However, they predict
that ISAW and ISAT are in different universality classes.
Importantly, while the upper critical dimension for ISAW is expected
to be $d_u=3$, the Shapir-Oono field theory gives $d_u = 4$ for
ISAT. Therefore, this implies generically that logarithmic corrections
occur at the $\theta$-point in four dimensions, which presumably
should occur at a finite temperature.  On the other hand the above
discussion about the hypothesised equivalence of the critical
phenomena of self-avoiding walks and the Edwards model may lead one to
the opposite conclusion, namely that the collapse transitions of ISAT,
ISAW and the (3-parameter) Edwards model lie in the same universality
class. Computer simulations in two and three dimensions have given
conflicting results and so it is of interest to ascertain whether
computer simulations in four dimensions may shed light on the question
of whether ISAW and ISAT share the same collapse universality class.

Recent developments have been two-fold. Firstly, preliminary evidence
of collapse in any polymer model in four dimensions was presented in
\cite{owczarek1998a-:a} using the so-called kinetic growth trail which
effectively simulates ISAT with a particular fixed set of
non-standard Boltzmann weights (and so fixed temperature). It was
argued from the results of the simulations that this set of  Boltzmann
weights corresponded to a so-called $\theta$-point where the ISAT
behaves in a way predicted, by the mean-field theory of the polymer
collapse, to be precisely at the collapse transition
point. Plausibility arguments then deduced the existence of the
collapse transition as the temperature is varied through this value.
Secondly, and more startling have been the results of simulations
\cite{owczarek2000a-:a,prellberg2000a-:a} of interacting self-avoiding 
walks in four dimensions. A dramatic collapse transition was located
in this model using simulations up to length $32678$. The nature of
this transition was described as a \emph{pseudo-first order}
transition because at any finite length the distribution of the
internal energy was bimodal when the temperature was near that at
which the specific heat attained its maximum (one indicator of the
transition's position), while it was argued that the effective latent
heat would slowly decay to zero in the thermodynamic limit (as the
length diverges) with an anomalous exponent. On the other hand the
specific heat seemed to diverge as the length was increased. Despite
these peculiar findings the simulational results were able to be
interpreted in terms of a framework explained some time ago by
Khokhlov
\cite{khokhlov1981a-a}, who applied the non-standard mean-field
approach of Lifshitz, Grosberg and Khokhlov (LGK)
\cite{lifshitz1968a-a,lifshitz1976a-a,lifshitz1978a-a} to arbitrary
dimensions.  This theory is based on a phenomenological free energy in
which the competition between a bulk free energy of a dense globule
and its surface tension drive the transition. The consequences of this
surface free energy were largely ignored in the polymer literature
until recently, when its effect on the scaling form of the finite-size
partition function was proposed and confirmed
\cite{owczarek1993c-:a,owczarek1993d-:a,grassberger1995a-a,nidras1996a-a}.
While markedly different for finite lengths this theory still predicts
a standard Gaussian $\theta$-point and in the infinite length limit a
second order phase transition with a jump in the specific heat (no
divergence as the temperature is varied: the divergence as length is
varied exists but ``disappears'' in the thermodynamic limit).

In this work we have simulated the standard ISAT model on the
four-dimensional hypercubic lattice over a range of temperatures using
the PERM algorithm \cite{grassberger1997a-a} in a similar fashion to
simulations implemented for ISAW \cite{owczarek2000a-:a}. This
algorithm is particularly efficient in high dimensions and so is well
suited to these simulations. We find evidence that parallels the findings
in the ISAW simulations \cite{owczarek2000a-:a} for a collapse
transition with the signature of pseudo-first order type. The
transition is, if anything more pronounced, with an equivalent
strength at shorter lengths. We demonstrate that our data is at least
as, if not more, consistent with the LGK theory as is the ISAW results
\cite{owczarek2000a-:a}. Hence we deduce that the collapse transitions
of ISAW and ISAT in four dimensions are in the same universality
class. It follows from these conclusions that the upper critical
dimension for collapse in ISAT is $d_u< 4$, and that possibly $d_u=3$ in
accord with the standard polymer theory and with the logarithmically
modified Gaussian state found in three-dimensional kinetic growth
trail simulations \cite{prellberg1995b-:a}.

\section{The ISAT model and scaling theories} 
\subsection{The ISAT model}
We define the ISAT model on the four-dimensional hypercubic lattice in
the following way. The lattice has coordination number $8$ and we
consider configurations $\ph_N$ of trails, or bond-avoiding walks, of
length $N$ (bonds) starting from a fixed origin. Let $m_k, k = 1,
\ldots ,4$ be the number of sites of the lattice that has been visited
$k$ times by the trail so that $\sum k\, m_k=N+1$.  The partition
function of a very general interacting trail model is
\begin{equation}
Z_N(\w_2,\w_3 ,\w_4) = \sum_{\ph_N} \w_2^{m_2} \w_3^{m_3} \w_4^{m_4}\; ,
\end{equation}	 
where $\w_k$ is the Boltzmann weight associated with $k$-visited
sites. The canonical model is one where every segment of the trail at
some contact site interacts with every other segment at that site, so
that
\begin{equation}
\w_k=\w^{k\choose 2}\mbox{ for } k = 2,3,4\; ,
\end{equation}
with $\w \equiv \w_2$.  This implies that in our specific case 
\begin{equation}
\w_2 = \w, \qquad \omega_3=\w^3, \qquad \omega_4=\w^6.
\end{equation}
The Boltzmann weight $\w=e^{\beta\epsilon}$ is associated with a
energy of contact $-\epsilon$ so that $\omega > 1.0$ for attractive
interactions. As we are only interested in the attractive case here
one can set $\epsilon=1$ for convenience. We define a reduced
finite-size free energy per step $\kappa_N(\omega)$ as
\begin{equation}
\kappa_N(\omega)={1\over N}\log Z_N(\omega).
\end{equation}
The usual free energy is related to this by $-\beta F_N\equiv
N\kappa_N(\omega)$. 

The average of any quantity $Q$ over the
ensemble set of allowed paths $\Omega_N$ of length $N$ is given
generically by
\begin{equation}
\langle Q \rangle_N (\omega)= \frac{\sum_{\varphi\in\Omega_N}
Q(\varphi) \w^{m(\varphi)}}{\sum_{\varphi\in\Omega_N} \w^{m(\varphi)}}\; 
\end{equation}
where $m=m_2+3m_3+6m_4$.
We define a normalised finite-size internal energy per step by
\begin{equation}
U_N(\omega) = \frac{\langle m \rangle}{N}\, ,
\end{equation}
and a normalised finite-size specific heat per step by
\begin{equation}
C_N(\omega) = \frac{\langle m^2 \rangle - \langle m \rangle^2}{N}\; .
\end{equation}
These quantities are related in the usual way to the reduced free
energy via $U_N=\partial\kappa_N/\partial\log\omega$ and $C_N=\partial
U_N/\partial\log\omega$.  Note that because of the absent factors of
temperature both $U_N$ and $C_N$ can take on non-zero values for
$\w=1.0$.

The thermodynamic limit in this problem is given by the limit 
$N\rightarrow\infty$ so that the thermodynamic free energy per 
step $f_{\infty}(\w)$ is given by
\begin{equation}
-\beta f_{\infty}(\w)= \kappa_\infty(\omega)=\lim_{N \rightarrow \infty}\kappa_N(\omega) \; .
\end{equation}
This quantity determines the partition function asymptotics, i.e. $Z_N(\omega)$ grows to
leading order exponentially as $\mu(\omega)^N$ with $\mu(\omega)=e^{\kappa_\infty(\omega)}$.

In our simulations we calculated two measures of the polymer's average
size.  Firstly, specifying a trail by the sequence of position vectors
${{\bf r}_0, {\bf r}_1, ..., {\bf r}_N}$ the average mean-square
end-to-end distance is
\begin{equation}
\langle R^2_e\rangle_N = \langle ({\bf r}_N- {\bf r}_0) \cdot ({\bf r}_N- {\bf r}_0) \rangle\; .
\end{equation} 
We shall use the symbol $R^2_{e,N}$ to be equivalent to 
\begin{equation}
R^2_{e,N} (\omega)\equiv \langle R^2_e\rangle_N .
\end{equation} 
The mean-square distance of a site occupied by the trail to the
endpoint, $\vecr_0$, is given by
\begin{equation}
\langle R_{m}^2 \rangle_N = \frac{1}{N+1}\sum_{i=0}^{N}\langle (\vecr_i - \vecr_0)\cdot(\vecr_i - \vecr_0)
\rangle\; .
\end{equation}
Again we define
\begin{equation}
R^2_{m,N}(\omega) \equiv \langle R^2_m\rangle_N.
\end{equation}
We also define the ratio 
\begin{equation}
B_N(\omega) = \frac{R^2_{m,N}}{R^2_{e,N}},
\end{equation}
which should have a universal limit in each
critical phase of the model.

\subsection{Standard polymer scaling theory}

Let us expand our view in this section to general polymer scaling
theory.  We first \emph{assume} that there is a single collapse
transition at some value of temperature and explore the
four-dimensional behaviour we might expect from the above defined
quantities in each of the phases.  The basic physics of the
coil-globule (collapse) transition can be understood by the
consideration of the average size of the polymer, $R_N$, either
$R_{e,N}$ or $R_{m,N}$, as a function of length $N$ in each of the
phases, so let us consider this first. 
Let us define an effective Boltzmann
weight $\omega=e^\beta$---of course, in any particular \emph{model} or
physical system the associated potentials may be different. 
Generally one always expects
that
\begin{equation}
R^2_{N} \sim a(\omega)\: N^{2\nu}\quad\mbox{as}\quad N\rightarrow\infty
\end{equation}
for any fixed value of $\omega$. 
In
four dimensions at infinite temperature, $\omega=1$, it has been
predicted
\cite{gennes1979a-a} that
\begin{equation}
\label{swollen-size}
R^2_{N} \sim a^{+}\: N \left(\log(N)\right)^{1/4} .
\end{equation}
If there does exist a collapse transition then one would expect that
this scaling extends (with a constant $a^{+}$ that depends on
temperature) down to the transition point. In the collapsed phase the
polymer is expected to assume a dense configuration on average and
hence the globular value of the radius-of-gyration exponent is
$\nu_g=1/d=1/4$ \cite{gennes1975a-a} with
\begin{equation}
\label{collapse-size}
R^2_{N} \sim a^{-}(\omega)\: N^{1/2}.
\end{equation}
Finally at some finite
transition temperature $1.0 < \omega_t < \infty$ a Gaussian scaling of
the radius of gyration should occur, that is
\begin{equation}
\label{theta-size}
R^2_{N} \sim a^\theta\: N,
\end{equation}
so that $\nu_t=1/2$. This Gaussian scaling is often used
(theoretically at least) to define the $\theta$-point
$\omega=\omega_\theta$ of an isolated polymer so that
$\omega_t=\omega_\theta$. The universal ratio $B_N$ is expected to
converge to the value $B_\infty= 1/2$ both in the swollen phase and at
$\omega_\theta$. However, one would expect slow logarithmic
corrections for $\omega < \omega_\theta$ and algebraic corrections at
$\omega_\theta$. For $\omega > \omega_\theta$ the phase is no longer
expected to be critical and so $B_\infty$ is no longer universal and
may be a non-constant function of $\omega$.

One can also consider the scaling of the partition function in each of
the regimes, given that there is a transition. For high temperatures
$1.0 < \omega < \omega_\theta$ one expects the infinite temperature
behaviour, which is
\cite{gennes1979a-a}
\begin{equation}
\label{swollen-partition-function}
Z_N \sim b^{+}(\omega)\: \mu(\omega)^N \:\left(\log N\right)^{1/4},
\end{equation}
while at low temperatures \cite{owczarek1993c-:a} one expects
asymptotics of the form
\begin{equation}
\label{collapse-partition-function}
Z_N \sim b^{-}(\omega)\: \mu(\omega)^N\: \mu_s(\omega)^{N^{3/4}}\, N^g
\end{equation}
where $\mu_s$ is related to the surface free energy of the polymer
globule and the exponent $g$ need not be universal (we only write it
for completeness of the asymptotic form).  For $\omega=\omega_\theta$
one expects
\begin{equation}
\label{theta-partition-function}
Z_N \sim b^{\theta}\: \mu(\omega_\theta)^N
\end{equation}
as a reflection of Gaussian behaviour.

In the thermodynamic limit the thermodynamic functions
$f_{\infty}(\omega)$, $U_{\infty}(\omega)$ and $C_{\infty}(\omega)$
are all expected to be analytic functions of $\omega$ except at
$\omega_\theta$. By using the correspondence to the tricritical model
\cite{gennes1975a-a} the mean field theory would imply that the
specific heat had a jump discontinuity at $\omega_\theta$ (the
associated exponent $\alpha=0$).  Of course, for finite $N$ there is
no sharp transition for an isolated polymer (unless one examines a
macroscopic number of such polymers).

\subsection{LGK theory}

We now provide a brief review of the predictions of the theory of
Lifshitz, Grosberg and Khokhlov (LGK) \cite{lifshitz1978a-a} as applied to
four-dimensional polymer collapse by Khokhlov
\cite{khokhlov1981a-a}. Firstly,
there exists a state where the excluded volume property of long chain
molecules is exactly cancelled by the attractive interactions between
parts of the polymer as mediated by the solvent. This is the
$\theta$-state. Secondly, when the attraction becomes even stronger
there eventuates a globular state where the polymer behaves as a
liquid drop. The results of the theory are based on a phenomenological
free energy of that globular state relative to the free energy of the
pure Gaussian state of the $\theta$-point at $T_\theta$. Hence the
condition applied to find the finite-size position of the transition
is to equate the relative free energy to zero. The relative free
energy is given as a sum of bulk and surface contributions which are,
in turn, given in terms of the second and third virial coefficients,
the length of the chains, and the linear size of the polymer found
from the globular density. In particular both the bulk and surface
free energies are proportional to the square of the second virial
coefficient.  It is assumed that on approaching the $\theta$-point the
second virial coefficient goes to zero linearly with temperature while
the third virial coefficient remains non-zero. Note that this implies
a quadratic dependence of the bulk free energy on the distance to the
$\theta$-point. Since the free energy has exponent $2-\alpha$ this
implies an exponent $\alpha=0$ (assuming that this part of the free
energy is singular). Therefore a second-order phase transition occurs
in the thermodynamic limit.

It is further assumed that the density in the globule is proportional
to the second virial coefficient and hence also goes to zero linearly
with temperature on approaching the $\theta$-point ($\beta=1$). Again
using an effective Boltzmann weight $\omega=e^{\beta}$ and defining
the transition as when the free energy is zero, Khokhlov
\cite{khokhlov1981a-a} finds a finite-size transition
temperature\footnote{for the sake of ease of expression in this
section we will use the word ``temperature'' to mean the effective
Boltzmann weight} $\omega_{c,N}$ which approaches the $\theta$-temperature
($\omega_\theta$) as
\begin{equation}
\label{shift}
\omega_{c,N} -\omega_\theta  \sim \frac{s}{N^{1/3}}
\end{equation}
for some constant $s$. That is, the polymer collapse \emph{shift}
exponent is $1/3$.  The width of the transition region $\Delta \omega$
at finite $N$ can be found from the free energy rewritten in terms of
this transition temperature to scale as
\begin{equation}
\label{width}
\Delta \omega \sim \frac{w}{N^{2/3}}
\end{equation}
for some constant $w$. That is, the polymer collapse \emph{crossover}
exponent is $2/3$. Hence note that the size of the crossover region is
asymptotically small relative to the shift of the transition.

%

Following the work \cite{prellberg2000a-:a} of Lifshitz, Grosberg and
Khokhlov \cite{lifshitz1978a-a} one can also calculate the change in
the internal energy over the crossover width of the transition $\Delta
\w$ as the latent heat (or ``heat of the transition'') by using
expression of the free energy in terms of the transition temperature as
\begin{equation}
\label{latent-heat}
\Delta U \sim  \frac{u^c}{N^{1/3}}.
\end{equation}
The corresponding height of the peak in the specific heat is
\begin{equation}
\label{specific-heat-peak}
C_{N}(\w_{c,N}) \sim h^c\; N^{1/3}.
\end{equation}

So to summarise the LGK picture, the theory predicts a thermodynamic
second-order transition at a Gaussian $\theta$-point with a jump in
the specific heat. For finite polymer length this transition is
shifted below the $\theta$-point by a temperature of the order of
$O(N^{-1/3})$ with the width of the transition of the order of
$O(N^{-2/3})$. Over this width there is a rapid change in the internal
energy that scales as $O(N^{-1/3})$: the important point here of
course is that this tends to zero for infinite length so the effect of
the peak in the specific heat is scaled away for $N$ large, leaving a
finite jump in the thermodynamic limit. To understand this further let
us consider the distribution of internal energy as a function of
temperature and length. For any $\omega$ below $\omega_\theta$ and
well above $\omega_{c,N}$ one expects the distribution of internal
energy to look like a single peaked distribution centred close to the
thermodynamic limit value: a Gaussian distribution is expected around
the peak with variance $O(N^{-1/2})$. In fact, this picture should be
valid for all temperatures outside the range $[\w_{c,N} -
O(N^{-2/3}),\w_{c,N} + O(N^{-2/3})]$. When we enter this region we
expect to see a double peaked distribution as in a first-order
transition region. For any temperature in this region there should be
two peaks in the internal energy distribution separated by a gap
$\delta U$ of the order of $\delta U\approx \Delta U \propto
O(N^{-1/3})$. Each peak is of Gaussian type with individual variances
again of the order of $O(N^{-1/2})$. Hence as $N$ increases the peaks
will become more and more distinct and relatively sharper but the peak
positions will be getting closer together. We refer to this scenario
as a \emph{pseudo}-first-order transition or, more correctly, as
first-order-like finite-size corrections to a second-order phase
transition. If there were a real first-order transition then the
distance between the peaks should converge to a non-zero constant. On
the other hand the transition is not a conventional second-order phase
transition with a well defined limit distribution of the internal
energy that is simply bimodal.


\section{Simulational results and analysis} 

We have simulated ISAT on a 4-dimensional hyper-cubic lattice using
the Pruned-Enriched Rosenbluth Method (PERM), a clever generalisation
of a simple kinetic growth algorithm
\cite{grassberger1997a-a,frauenkron1998a-a}.  PERM builds upon the
Rosenbluth-Rosenbluth method \cite{rosenbluth1955a-a}, in which trails
are generated by simply growing an existing trail kinetically but
overcomes the exponential ``attrition'' and re-weighting needed in
this approach by a combination of enrichment and pruning strategies. 
Our implementation here follows our previous ISAW work
\cite{prellberg2000a-:a,owczarek2000a-:a}. Briefly,  we chose upper
and lower thresholds $W^u$ and $W^l$, for enrichment and pruning
respectively, proportional to the current estimate of the average
weight of a trail at length $N$, $\langle Z_N\rangle/s_N$, where $s_N$
is the number of generated samples at length $N$, and $\langle
Z_N\rangle$ is the current estimate of the partition function at
length $N$. That is to say, $W^u_N=c^u_N\langle Z_N\rangle/s_N$, $
W^l_N=c^l_N\langle Z_N\rangle/s_N$.  In order to enforce an even
sample size distribution we allowed for dynamic adjustment of $c^u_N$
and $c^l_N$, keeping the quotient of the thresholds $Q=c^u_N/c^l_N$
constant. To stabilise the dynamic adjustment, we enforced
$c^u_N>c^u_{min}$ and $c^l_N<c^l_{max}$. As in the ISAW work, we chose
$c^u_{min}=2$ and $c^l_{max}=1/2$. For each run, we attempted to
choose the smallest threshold quotient $Q$ for which we could obtain
an even sample size distribution.

Each run had a maximum length $N_{max}$ set and while individual runs
gave information about shorter lengths we collected data from
independent runs at some shorter lengths to guarantee statistical
independence.  Simulations were conducted with the maximum lengths
$N_{max}$ set to 512, 1024, 2048, and 4096, with values of $\omega$
ranging from $1.0$ to $2.07$ for $N_{max}=512$, from $1.0$ to $2.00$
for $N_{max}=1024$, from $1.0$ to $1.78$ for $N_{max}=2048$, from
$1.0$ to $1.67$ for $N_{max}=4096$. We also ran many closer spaced
simulations in the range of $\omega$ from $1.4$ to $1.42$ at length
$N_{max}=16384$.  At each fixed $\omega$, we generated at least $10^7$
trails. To illustrate the computational effort, the generation of a
sample of size $10^7$ at length $N_{max}=16384$ took about $2$ weeks
CPU time on a $600$ MHz DEC Alpha. The threshold quotient $Q$ used ranged
from $10$ to $80$ with larger values of $Q$ needed for higher
$\omega$.  We also performed one large-scale simulation deep in the
collapsed regime, with $N_{max}=512$ and $\omega=4.0$, for which it
was necessary to increase the threshold quotient up to $Q=1000$.

We computed statistics for $R_{e,N}^2$ and $R_{m,N}^2$, the partition
function $Z_N$, the internal energy $U_N$ and specific heat
$C_N$. Moreover, we generated the distribution of the number of
interactions at $N_{max}$. The distributions obtained at various
temperatures were then combined using the multiple histogram method
\cite{ferrenberg1988a-a}.

The disadvantage of PERM is that due to the enrichment the generated
data is not independent. All the data generated during one ``tour'',
i.e. between two successive returns of the algorithm to length $0$, is
correlated. Therefore, we kept track of the statistics of tour sizes
$t$ to get a rough idea of the quality of the data. In our statistical
evaluation we use (somewhat arbitrarily) the quotient of $s_N$ and
$\sqrt{\langle t^2\rangle}$ as a measure of an effective independent
sample size. This is correct as long as the tour sizes don't fluctuate
too strongly, and, more importantly, as long as individual tours
explore the sample space evenly. When simulating in the collapsed
phase, both of these assumptions break down, and the sample is
dominated by few huge tours. Moreover, the pruning and enrichment
rates become so large that the efficiency of the algorithm is
significantly decreased. Error bars are only given for high
temperature and $\theta$-point simulation figures  (figures \ref{figure1} and
\ref{figure3})  and are based on the method
described above, although we always computed error estimates.  No
error bars are given in the rest of the figures because of the
performance of the algorithm in and near the collapsed phase, even
though the data seemed converged sufficiently.

Let us first discuss the scaling of the mean-squared end-to-end
distance normalised by trail length, $R_{e,N}^2/N$ and the
mean-squared distance of a site occupied by the trail to the trail's
end-point, $R_{m,N}^2/N$. In the swollen phase, our results are in
correspondence with the logarithmic corrections seen by Grassberger
\cite{grassberger1994a-a}. As in that paper, we observe that
$R_{e,N}^2$ grows faster than $N$ for $\omega$ near $1.0$, and fitting
to $N(\log N)^c$ at $\omega=1.0$ gives an effective exponent close to
that predicted by field theory ($1/4$). This value shifts as $\w$ is
increased indicating the presence of strong temperature-dependent
correction terms.

By considering when the quantity $R_{e,N}^2/N$ approaches a constant
we narrowed our search for the $\theta$-point to the region $\w=1.40$
to $\w=1.42$. In this region we extended our simulations to trails of
length $16384$. Figure \ref{figure1} shows a plot of $R_{e,N}^2/N$
versus $1/N$ for values of $\omega$ between $1.408$ and $1.420$. At
$\omega=1.414(3)$ we have an approximate linear asymptotic dependence
of $R_{e,N}^2$ on $N$. Moreover, at $\omega=1.414$ we estimate from
our data $B_N=R_{m,N}^2/R_{e,N}^2=0.5000(2)$, which is also indicative
of Gaussian behaviour: the precision of this estimate stems from the
weakness of the corrections to scaling at this point.

As shown in Figure \ref{figure2} for $\omega=4.0$, $R_{e,N}^2$ changes
non-monotonically in $N$! After an initial increase, the size of the
polymer actually starts to shrink around $N=50$ as it undergoes
collapse corresponding to a rapid increase of the density. For large
enough $N$, we expect to see the true collapsed behaviour, i.e.\
$R_{e,N}^2$ growing again as $N^{1/2}$, but while we see $R_{e,N}^2$
just starting to increase again, the asymptotic regime is beyond the
reach of our PERM simulations on current computer hardware.

Let us now discuss the scaling of the partition function. The swollen
phase and the $\theta$-point behaviour can also be clearly identified
from the free-energy scaling. In the swollen phase we find again the
same behaviour as \cite{grassberger1994a-a}.  The presence of
logarithmic corrections is consistent with our data. At $\omega=1$, we
estimate $\mu(1)=\mu_{SAT}=6.926080(2)$.  In the $\theta$-region, an
analysis shows that here $Z_N$ scales as $\mu^N$ with weak $1/N$
corrections.  Figure \ref{figure3} shows $Z_N/Z_{N/2}^2$ plotted
versus $1/N$ from which we estimate the $\theta$-point to be
$\omega_\theta=1.414(3)$ and $\mu_\theta=7.0016(6)$. (At fixed
$\omega$, the accuracy is of course higher: for $\omega=1.414$, we
estimate $\mu=7.0015714(5)$.) In the collapsed region, one expects the
finite-size free energy to have a strong correction term of the order
$N^{-1/4}$ due to surface effects. Figure \ref{figure4} shows this for
$\omega=4.0$. As argued above, the globule starts to collapse when the
length is above $N=50$, and we notice here the onset of a
corresponding strong change in the behaviour of the finite-size free
energy around this length ($N^{-1/4}\approx0.35$). Even though we
cannot simulate long enough chain lengths to clearly determine the
precise nature of the correction term, our data is certainly
compatible with a $N^{-1/4}$ correction for $N^{-1/4}<0.3$ (i.e.\ $N >
150$).

In order to study the collapse transition more closely, we now focus
our attention on the internal energy and specific heat. As can be seen
from Figure \ref{figure5}, the specific heat has a sharply peaked
graph for each length that becomes more sharply peaked as $N$
increases. The transition region becomes sharper and stays well
separated from the $\theta$-point, even though the location of the
transition (peak in the specific heat) approaches the $\theta$-point
slowly.
The scaling of the shift of the transition towards the $\theta$-point,
$\omega_{c,N}-\omega_\theta$, and the sharpening of the transition
width, $\Delta\omega$, are both shown in Figure \ref{figure6}. Here, we
defined the location of the collapse transition by the location of the
specific heat peak, and the width of the transition is given by the
interval in which the specific heat is greater or equal to half the
value of the peak height. Expecting from the LGK theory that
$\omega_{c,N}-\omega_\theta$ scales as $N^{-1/3}$ and that
$\Delta\omega$ scales as $N^{-2/3}$, we plot both
$N^{1/3}(\omega_{c,N}-\omega_\theta)$ and $\Delta\omega N^{2/3}$
versus $N^{-2/3}$ which was chosen empirically. Both quantities can be
seen to be asymptotic to constants: on the graph extrapolations give
non-zero intercepts. Hence, Figure
\ref{figure6} shows that the LGK predictions are compatible with our
simulations. We do note that the corrections to scaling for
$\Delta\omega$ are much larger than for
$\omega_{c,N}-\omega_\theta$.

The character of the transition becomes apparent if one plots the
internal energy density distribution (rescaled density of
interactions) at the finite-size collapse transition,
$\omega_{c,N}$. Figure \ref{figure7} shows the emergence of a bimodal
distribution.  At length $512$ one sees a slight non-convexity, which
at length $4096$ has evolved into a distribution dominated by two
sharp and well-separated peaks. The values of the minima and maxima of
the distribution are different by two orders of magnitude. This
bimodal distribution means that as $\omega$ is increased through the
transition region the density distribution switches from the peak
located at a small value of contacts to the peak located at a larger
value of contacts, corresponding to a sudden change in the internal
energy.  In the collapsed phase, the width of the peak is much wider
than in the swollen phase, implying a larger specific heat. It is this
difference between the swollen and collapsed phases' specific heats
that will eventually become the thermodynamic second order jump. The
rapid first-order like switch between two peaks in the distribution
becomes more pronounced at larger polymer lengths since the depth of
the ``valley'' between the two peaks becomes relatively larger.

Continuing with the scaling predictions from Khoklov theory, a suitably
defined finite-size latent heat, $\Delta Q$, should tend to zero as
$N^{-1/3}$ in the thermodynamic limit. One possible measure of this
latent heat is given by the product of specific heat peak
$C_N(\omega_{c,N})$ and specific heat width $\Delta\omega$, and
another is given by the distance $\delta U$ of the peaks in the
bimodal internal energy distribution.  Figure \ref{figure8} shows the
behaviour of both of these quantities.  One notices two things from
this figure. Firstly, it indicates that $C_N(\omega_{c,N})\Delta\omega$
decreases to zero linearly in $N^{-1/3}$ as predicted. However, even at length
$N=2048$ ($N^{-1/3}\approx 0.08$) there is considerable discrepancy
between the two quantities plotted, so that one needs to be cautious in the
interpretation of the scaling behaviour.  The explanation for the
discrepancy between the two quantities  is of course that in order
to observe the asymptotic behaviour the two peaks in the histogram
have to be well separated and distinct, and that this is only really the case
when $N$ is of the order of $10^3$. We caution that Figure
\ref{figure8} alone is not sufficient to discriminate between the scenario
proposed here and a real first-order transition in the thermodynamic limit, 
but we believe the rest of our data and other theoretical facts provide a more
consistent picture.

When comparing our data with the simulations for ISAW
\cite{prellberg2000a-:a} we note further that the bimodal distribution
emerges for ISAT at much shorter configurations, so that the peaks in
the distribution for ISAT at length $N=512$ are already more
pronounced than the peaks in the distribution for ISAW at length
$N=2048$.  To quantify this observation, we turn to the scaling
predictions of LGK theory. An important parameter in the theory is the
quotient $a^d/v$, where $a$ is the mean-square distance between two
subsequent monomers (repeated unit element of the polymer: equivalent
to occupied sites of the lattice model) along a chain and $v$ is the
effective excluded volume of a monomer, defined via the vanishing of
the second virial coefficient at the $\theta$-temperature. For instance, 
the shift of the transition temperature (cf. equation
\ref{shift}) is given more explicitly by
\begin{equation}
{\omega_{c,N}-\omega_\theta\over\omega_\theta}\sim\left({\tilde sa^4\over Nv}\right)^{1/3}
\end{equation}
where $\tilde s$ is a constant proportional to the quotient of the third 
virial coeffficient and the excluded volume squared.
From Figure \ref{figure6} we estimate that
$N^{1/3}(\omega_{c,N}-\omega_\theta)$ asymptotes to $3.4(1)$ for ISAT,
and for ISAW we estimate for the same quantity the value $0.92(3)$
\cite{prellberg2000a-:a}. Identifying $a$ with the lattice constant, 
which in both models is set equal to one, we can get a rough estimate for 
the relative size of the effective excluded volume $v$ in both models. 
We obtain
\begin{equation}
{v_{SAT}\over v_{SAW}}\approx 0.03 {\tilde s_{SAT}\over\tilde s_{SAW}}
\end{equation}
and thereby quantify the intuitive notion that the excluded volume
effect is numerically ``weaker'' in trails than in walks, though of
the same basic type.

In conclusion, our ISAT simulations elucidate further the structure of the
polymer collapse transition in four dimensions.  We show conclusively
that there is indeed a collapse transition at a finite
temperature. Secondly, we find evidence for a $\theta$-temperature at
which the polymer is well approximated by Gaussian behaviour as well
as for a collapse transition which is well separated from the
$\theta$-point. The collapse transition shows many first-order like
features, such as a bimodal distribution in the internal energy. An
analysis of the scaling behaviour of this transition in the context of
the theory of Lifshitz, Grosberg and Khokhlov
\cite{lifshitz1978a-a,khokhlov1981a-a} shows that a consistent
interpretation of these findings is that of first-order like
finite-size corrections to a thermodynamic second-order
transition. These finding are essentially the same as those made
recently for ISAW collapse in four dimensions
\cite{owczarek2000a-:a}. Consequently, we deduce that the upper
critical dimension for ISAT is $d_u < 4$ (most likely $3$) and not
$d_u=4$ as was previously predicted.

\section*{Acknowledgements} 
Financial support from the Australian Research Council is gratefully
acknowledged by ALO.


\newpage 

\begin{figure}
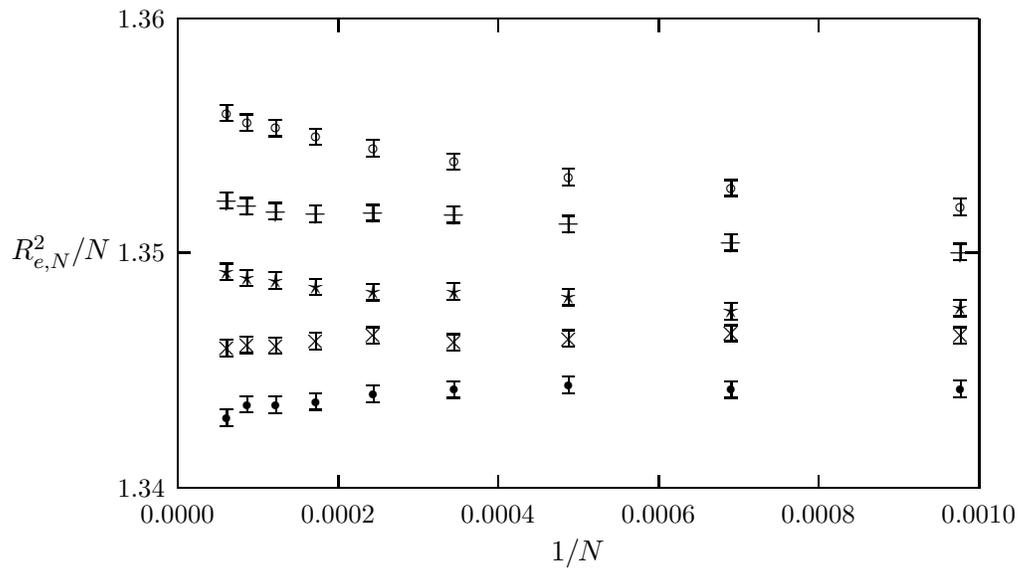

\caption{\it $R_{e,N}^2/N$ versus $1/N$ in the $\theta$-region:
$\omega=1.408$, $1.411$, $1.414$, $1.417$, $1.420$ from top to bottom.} 
\end{figure}

\begin{figure}
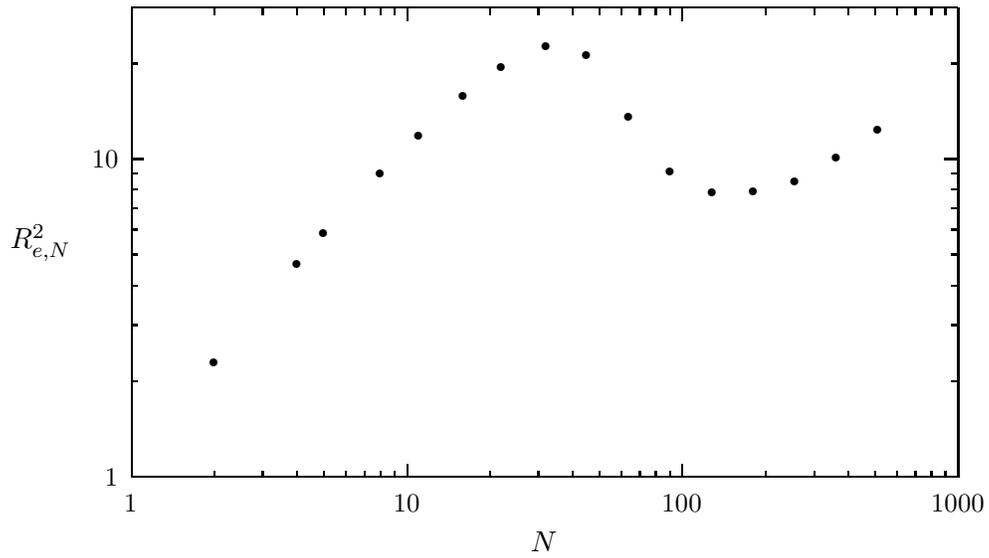

\caption{\it $R_{e,N}^2$ versus $N$ for $\omega=4.0$ up to length $512$.}
\end{figure}

\begin{figure}
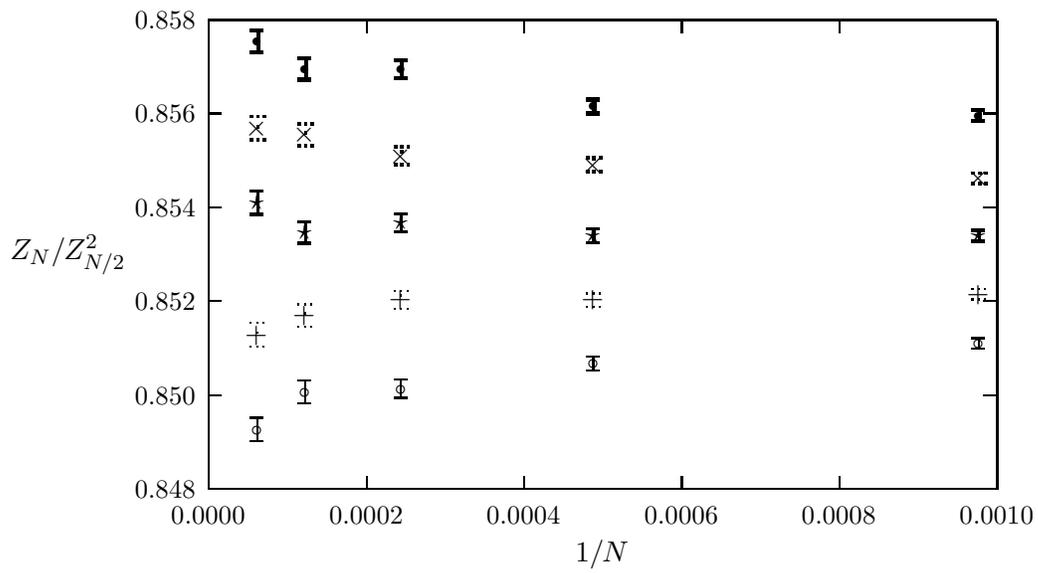

\caption{\it $Z_N/Z_{N/2}^2$ versus $1/N$ in the $\theta$-region:  
$\omega=1.408$, $1.411$, $1.414$, $1.417$, $1.420$ from bottom to top.
}
\end{figure}

\begin{figure}
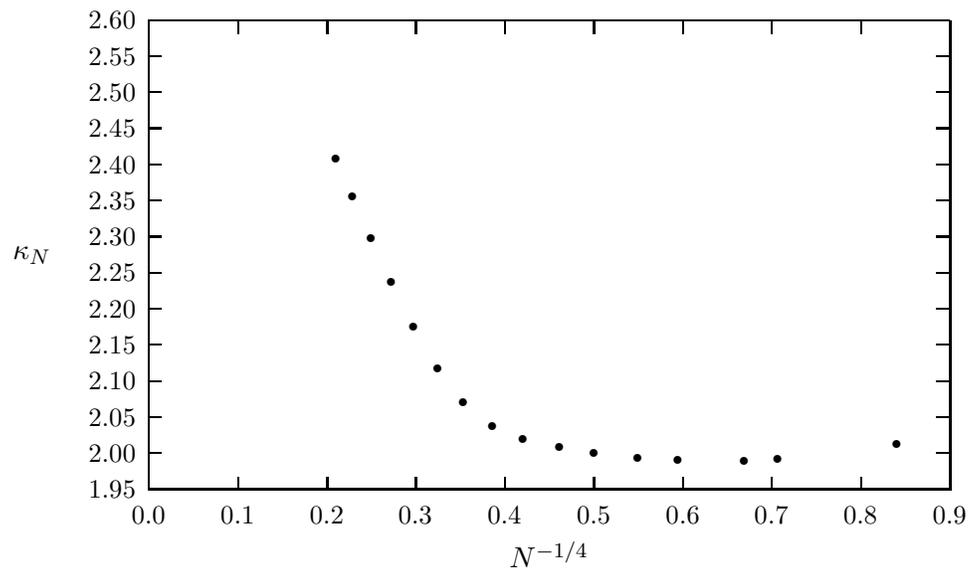

\caption{\it Finite-size free energy $\kappa_N$ versus $N^{-1/4}$ for $\omega=4.0$.}
\end{figure}

\begin{figure}
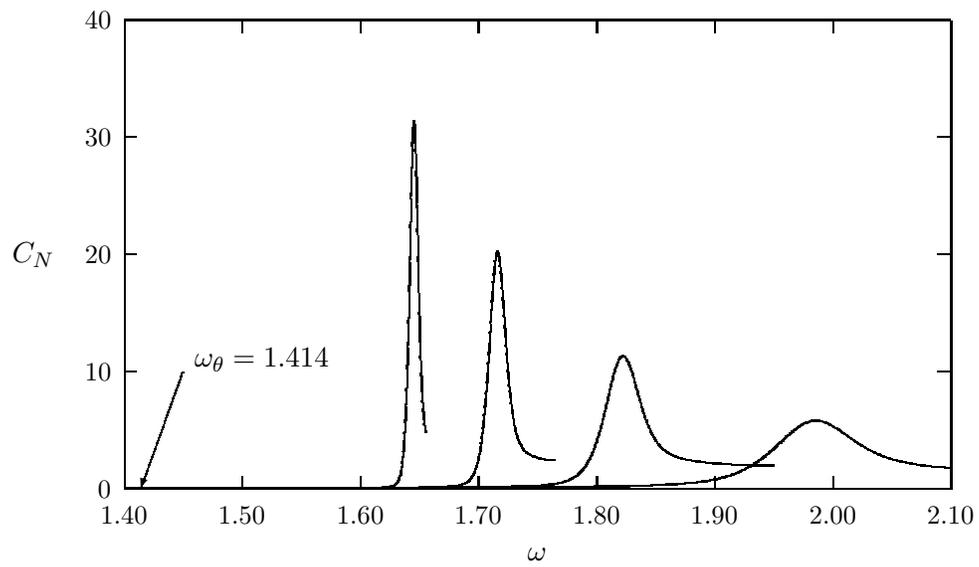

\caption{\it Specific heat $C_N$ versus $\omega$ for lengths
$512$, $1024$, $2048$, and $4096$  from right to left
respectively, using the multi-histogram method.} 
\end{figure}

\begin{figure}
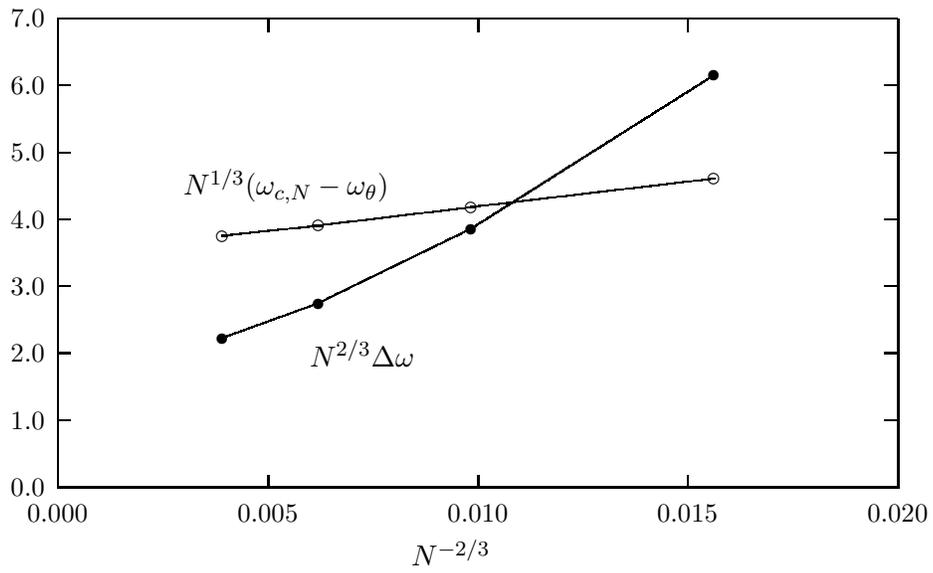

\caption{\it Scaling of the transition: shift and width of the
collapse region. Shown are the scaling combinations 
$N^{1/3}(\omega_{c,N}-\omega_\theta)$ and $N^{2/3}\Delta\omega$ versus $N^{-2/3}$.} 
\end{figure}

\begin{figure}
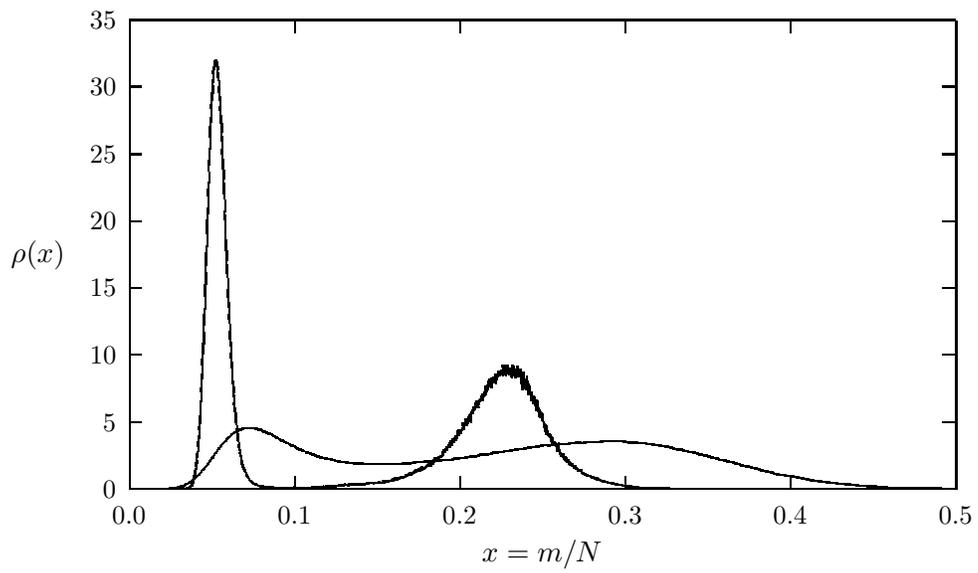

\caption{\it Internal energy density distributions at $\omega_{c,N}$
for $512$ and $4096$. The more highly peaked distribution
is associated with length $4096$.}
\end{figure}

\begin{figure}
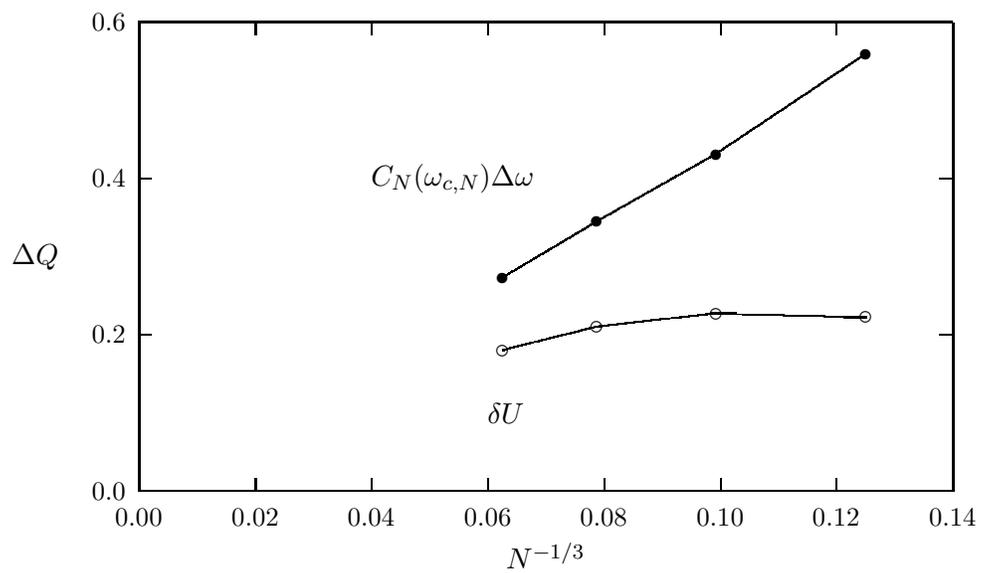

\caption{\it Scaling of the latent heat $\Delta U$:
our two measures of $\Delta U$, $C_N(\omega_{c,N})\Delta\omega$ and
peak distance $\delta U$ are plotted versus $N^{-1/3}$.}
\end{figure}

\clearpage
\newpage

\setcounter{figure}{0}

\begin{figure}
\begin{center}
\setlength{\unitlength}{0.240900pt}
\ifx\plotpoint\undefined\newsavebox{\plotpoint}\fi
\sbox{\plotpoint}{\rule[-0.200pt]{0.400pt}{0.400pt}}%

\end{center}
\caption{\it $R_{e,N}^2/N$ versus $1/N$ in the $\theta$-region:
$\omega=1.408$, $1.411$, $1.414$, $1.417$, $1.420$ from top to bottom.} 
\label{figure1} 
\end{figure}

\clearpage
\newpage

\begin{figure}
\begin{center}
\setlength{\unitlength}{0.240900pt}
\ifx\plotpoint\undefined\newsavebox{\plotpoint}\fi
\sbox{\plotpoint}{\rule[-0.200pt]{0.400pt}{0.400pt}}%
%
\end{center}
\caption{\it $R_{e,N}^2$ versus $N$ for $\omega=4.0$ up to length $512$.}
\label{figure2} 
\end{figure}

\clearpage
\newpage

\begin{figure}
\begin{center}
\setlength{\unitlength}{0.240900pt}
\ifx\plotpoint\undefined\newsavebox{\plotpoint}\fi
\sbox{\plotpoint}{\rule[-0.200pt]{0.400pt}{0.400pt}}%
%
\end{center}
\caption{\it $Z_N/Z_{N/2}^2$ versus $1/N$ in the $\theta$-region:  
$\omega=1.408$, $1.411$, $1.414$, $1.417$, $1.420$ from bottom to top.
}
\label{figure3} 
\end{figure}

\clearpage
\newpage

\begin{figure}
\begin{center}
\setlength{\unitlength}{0.240900pt}
\ifx\plotpoint\undefined\newsavebox{\plotpoint}\fi
\sbox{\plotpoint}{\rule[-0.200pt]{0.400pt}{0.400pt}}%
%
\end{center}
\caption{\it Finite-size free energy $\kappa_N$ versus $N^{-1/4}$ for $\omega=4.0$.}
\label{figure4} 
\end{figure}

\clearpage
\newpage

\begin{figure}
\begin{center}
\setlength{\unitlength}{0.240900pt}
\ifx\plotpoint\undefined\newsavebox{\plotpoint}\fi
\sbox{\plotpoint}{\rule[-0.200pt]{0.400pt}{0.400pt}}%
%
\end{center}
\caption{\it Specific heat $C_N$ versus $\omega$ for lengths
$512$, $1024$, $2048$, and $4096$  from right to left
respectively, using the multi-histogram method.} 
\label{figure5} 
\end{figure}

\clearpage
\newpage

\begin{figure}
\begin{center}
\setlength{\unitlength}{0.240900pt}
\ifx\plotpoint\undefined\newsavebox{\plotpoint}\fi
\sbox{\plotpoint}{\rule[-0.200pt]{0.400pt}{0.400pt}}%
%
\end{center}
\caption{\it Scaling of the transition: shift and width of the
collapse region. Shown are the scaling combinations 
$N^{1/3}(\omega_{c,N}-\omega_\theta)$ and $N^{2/3}\Delta\omega$ versus $N^{-2/3}$.} 
\label{figure6} 
\end{figure}

\clearpage
\newpage

\begin{figure}
\begin{center}
\setlength{\unitlength}{0.240900pt}
\ifx\plotpoint\undefined\newsavebox{\plotpoint}\fi
\sbox{\plotpoint}{\rule[-0.200pt]{0.400pt}{0.400pt}}%
%
\end{center}
\caption{\it Internal energy density distributions at $\omega_{c,N}$
for $512$ and $4096$. The more highly peaked distribution
is associated with length $4096$.}
\label{figure7} 
\end{figure}

\clearpage
\newpage

\begin{figure}
\begin{center}
\setlength{\unitlength}{0.240900pt}
\ifx\plotpoint\undefined\newsavebox{\plotpoint}\fi
\sbox{\plotpoint}{\rule[-0.200pt]{0.400pt}{0.400pt}}%
%
\end{center}
\caption{\it Scaling of the latent heat $\Delta U$:
our two measures of $\Delta U$, $C_N(\omega_{c,N})\Delta\omega$ and
peak distance $\delta U$ are plotted versus $N^{-1/3}$.}
\label{figure8} 
\end{figure}

\end{document}